\documentclass[12pt]{article}
\usepackage{amsmath}
\usepackage[psamsfonts]{amssymb}
\usepackage{graphicx}

\begin{document}

\footnotesize{\noindent {\it Hadronic Journal Supplement} {\bf
18}, no. 4, pp. 425-456 (2003)}

\begin{center}
\section*{ON THE NATURE OF THE ELECTRIC CHARGE}
\end{center}
\vspace{4mm}
\begin{center}
{\large {\bf Volodymyr Krasnoholovets}}
\end{center}

\begin{center}
{Institute for Basic Research \\ 90 East Winds Court, Palm Harbor,
FL 34683, U.S.A. \\
 Home Page http://www.inerton.kiev.ua}

\vspace{4mm} \hspace{15 cm} 8 April 2003

\end{center}

\vspace {2mm} {\small
\begin{center}
{\bf Abstract}
\end{center}

The geometry of the elementary charge is studied in the framework
of the concept of space considered as a tessellation lattice
('tessellattice'), which has recently been developed by M. Bounias
and the author. The descriptive-geometric sense of the electric
and magnetic fields and their carriers  -- photons -- is analyzed.
The notion of the scalar and vector potentials of a charged
particle and their behavior at the motion of the particle along
its path is investigated in detail. Based on the potentials, the
Lagrangian leading to the Maxwell equations is constructed. The
distinctive properties of the inerton and the photon -- two basic
elementary excitations, or quasi-particles of the space
tessellattice -- are discussed. Summarizing, we may say that the
work suggests the detailed interpretation of the Maxwell equations
in terms of the submicroscopic approach to Nature. \\

\bigskip

{\bf Key words:} space, tessellattice, electric charge, Maxwell
equations, photon, inerton

\bigskip

{\bf PACS:} 01.55+b General physics; 03.50De Classical
electromagnetism, Maxwell equations; 11.90+t Other topics in
general theory of fields and particles

\newpage

\begin{flushright} \footnotesize{The electric charge is kept and
\ \ \ \ \ \ \ \ \ \  \ \ \ \ \ \ \
\\   \ \ \quad\quad\quad\quad\quad\quad
 plays on the surface of a particle. \ \ \ \ \ \ \ \
 \quad\quad\quad   \\

\hspace{5.3cm} \break  $The$ $\d{\it R}gveda$, \ 1.28.9; \
9.65.25; \ 9.66.29 \quad \quad \quad}
\end{flushright}

\vspace{2mm}

\section*{\small 1. INTRODUCTION}

\hspace*{\parindent} A very abstract and formal nature of modern
schemes of geometrization of fundamental physical interactions is
a matter of common knowledge. Among such kinds  of approaches
there are schemes of grand unification of electromagnetic, weak
and strong interactions, theories of string, superstring and
supergravity, etc. In these theories the description of sources of
fields -- the mass and the electric charge -- is realized by means
of abstract classical or wave fields of an indeterminate nature
(but fundamental as are suggested!) with a respective statistical
interpretation. One points out here only some of the approaches
[1-8]. A mathematical model (field presentation) of a single
photon as a real finite object has been developed in papers [9].
We should point out an interesting model by Hofer [10] aimed at
the description of the electromagnetic field to so-called natural
units in the framework of which the electric field and the vector
potential change their dimensionality to typical mechanical units.
Recently, models of charged particles that are characterized by
appropriate electromagnetic clouds, Bagan {\it et al.} [11], or
clouds of soft photons, Greenberg [12], have been proposed; in
those models infrared divergences disappear just due to the
original dressing of particles. Todorov  [13] treated electrical
charges as something that followed from the gauge-invariant Dirac
and Klein-Gordon equations. Regarding the further investigation of
the Maxwell equations we may point out in particular an approach
by Drummond [14] aimed at interchanging the magnetic and electric
variables and the formation of the common complex variable.

However, it must be emphasized that the mentioned models and all
other ones of quantum electrodynamics determine photons and
consider their interaction with particles only in abstract phase
spaces such as the momentum one, but not in the real space.

    At the same time the structure of the real physical space
considered as a structured vacuum, a quantum aether, or a space
net on the background of which all physical processes take place
has come under the scrutiny of science in the last decade (see,
e.g. Refs. [15-20]). Specifically, in the author's works [19-23]
there have been developed a concept that makes it possible to look
at the notion of mass and the motion of a massive particle from
the other viewpoint that fundamentally differs from standard field
approaches. The further development has allowed the deriving the
macroscopic gravitation from submicroscopic first principles
[24,25].

It turned out that the author's approach was in excellent
agreement with the pure mathematical research of the construction
of space carried out by Bounias and Bonaly [26], and Bounias [27].
This allowed M. Bounias and the author joined forces in order to
develop founding principles about mathematical constitution of
space [28] and then to construct the physical space, i.e. to
derive matter and physics laws from the constitution of space and
combination rules permissible by the space [29]. In these works we
start from set theory, topology and fractal geometry and has shown
that the real space-time is characterized by a topologically
discrete structure. An abstract lattice of empty set cells
accounts for a primary substrate in the real physical space.
Space-time is represented by ordered sequences of topologically
closed Poincar$\acute{\rm e}$ sections of this primary space with
a lattice structure called the 'tessellattice'. The interaction of
a moving particle-like deformed cell with the surrounding
tessellattice involves a fractal decomposition process that
supports the existence and properties of previously postulated
inerton clouds as associated to particles.

We have shown [29,30] that if a deformation of primary cells
involves a fractal transformation to objects, there occurs an
alteration in the dimensionality of the cell(s) and the
homeomorphism is not conserved. In this event the fractal kernel
stands for a particle state and the reduction of the corresponding
volume of the cell is compensated by morphic changes of
surrounding cells. The fractality is specified by quanta and
combination rules [30] that determine oriented sequences with
continuity of set-distance functions providing a space-time-like
structure that facilitates the aggregation of the mentioned
deformations into fractal forms standing for massive objects. An
appropriate dilatation of space is induced outside the
aggregation. At a submicroscopic scale the families of fractal
deformations produce families of particle-like structures, but at
large scale an apparent expansion has to occur.

It has been argued [22] that the size of a primary cell of the
space tessellattice is probably on the order of $10^{-30}$ m (it
is the scale at which, according to the theory of grand
unification of interactions, all types of interactions should
coincide). In a flat, or degenerate space all cells, or
superparticles, are found in a state degenerated over all
multiplets and can be characterized by a mean volume ${\cal
V}_{\rm sup}$. The creation of a mass particle means the change in
the initial volume of the correlative superparticle, i.e.  ${\cal
V}_{\rm sup}$ passes to the other fixed value ${\cal V}_{\rm
part}$ and the rest mass of the particle is determined as
$M_{{\kern 1pt} 0} \propto {\cal V}_{\sup}/{\cal V}_{\rm part}$. A
detailed study on the generation of mass in a cell of the
tessellattice has been performed in Refs. [29,30] in which we have
shown that the mass of a particle is a function of the
fractal-related decrease of the volume of the cell,
 $$
M \propto \frac 1{{\cal V}_{\rm part}} {\kern 2pt}({\rm e}_{\kern
1pt \nu} -1)_{{\kern 1pt}{\rm e}_{{\kern 1pt}\nu \geq 1}}
 $$
where $(\rm e)$ is the Bouligand exponent, and $(\rm e - 1)$ is
the gain in dimensionality given by the fractal iteration. Just
the volume decrease is not sufficient for providing a cell with
mass, because a dimensional increase is a necessary condition.

As has been shown in Ref. [24], a moving particle constantly
exchanges by mass with the particle's inertons: the inert mass of
the particle $M_0/\sqrt{1-v^2_0/c^2}$ is periodically dissociated
to $M_0$  (on each odd segment of the particle path $\lambda/2$
where $\lambda$ is the particle's de Broglie wavelength) and then
the particle re-absorbs inertons (on each even segment $\lambda/2$
of the particle path) and its mass is restored to $M_0/
\sqrt{1-v_0^2/c^2}$. Thus gravitons of the general theory of
relativity (which are not a realistic solution [31,{\kern 0.5pt}
32]) give way to inertons.

    The existence of inertons was demonstrated in papers [32-34].
Because of that it is quite reasonable to attempt to extend the
author's concept and study how an electric charge appears in the
canonical elementary particle within the framework of the space
tessellattice and how the charge moves dressed by a cloud of
elementary excitations that carry both inert and electromagnetic
properties. Account must be taken of the fact that the motion of
the charge should be coordinated with all the aspects of
electrodynamics. And the first step on the road to such a theory
has already been made [35]: the photon allows an interpretation of
an excitation migrated in a discrete space, i.e. the
tessellattice.

It is well known that atoms in a crystal, in view of interaction,
are found in the uninterrupted collective motion. When this is the
case, each of atoms has as many non-equivalent wave vectors as
there are atoms in the crystal lattice (see, e.g. Ref. [36]). Let
us transfer this property to the space tessellattice, especially
as in Ref. [30] we have shown that topological structures
available at elementary scales can be distributed down to cosmic
scales. In particular, this is true for the topological notion of
the charge formed when a quantum of fractal deformations collapses
into one single cell: the concavity may be determined as the
negative charge and the convexity may be defined as the positive
charge.

Before going into the further development of topological details
pertinent to the charge construction and its behavior, we first of
all should construct a simple physical model of the charge in the
space tessellattice employing conventional mathematics. Let us try
to do that in the present work.

\section*{\small  2. COLLECTIVE BEHAVIOR OF  SUPERPARTICLES}

\vspace{2mm} \hspace*{\parindent} Let ${\mathfrak{N}}$ be the
quantity of cells (or superparticles, in other words) that the
whole space tessellattice contains. Of course the value of
${\mathfrak{N}}$ is inconceivable huge. Nonetheless it can be
roughly estimated. The size of the observed part of the universe
is on the order of $10^{26}$ m (see, e.g. Ref. [37]). The
superparticle size has been estimated as $10^{-30}$ m. Therefore,
${\mathfrak{N}} \sim (10^{26} {\rm m})^3/(10^{-30}{\rm m})^3=
10^{168}$.

   The availability of direct contact between neighboring superparticles
means that they are characterized by a coupling energy. Such kind
of a connection brings into existence collective vibrations of
superparticles and hence each of superparticles is influenced by
remained ${{\mathfrak{N}}}-1$. The vibration process lies in the
fact that centers of equilibrium positions of superparticles
vibrate with amplitudes $a_n$ ($\overline {n=1,
{{\mathfrak{N}}}/2}$) and each of the superparticles takes part in
${{\mathfrak{N}}}$ different vibrations simultaneously. When
superparticles are in contact, the volume ${\cal V}_{\rm sup}$ of
a degenerate superparticle should be smaller than the volume
${\cal V}_0$ of an absolutely free superparticle, which is reached
when the superparticle is removed from the space tessellattice.
Let the rest mass of the superparticle (i.e. degenerate
superparticle) in the tessellattice be defined as ${\cal M}_{\rm
sup} \propto {\cal V}_0/{\cal V}_{\rm sup}$. In this case $\Delta
{\cal M}_{0,{\rm sup}} \propto {\cal V}_0\times(1/{\cal V}_{\rm
sup} - 1/{\cal V}_0)$ can be called the superparticle's defect
mass produced by the coupling. To regard space as the zeroth
vacuum (in terms of quantum field theory), one needs to consider
the mass ${\cal M}_{\rm sup}$ as a frame of reference of all
massive excitations including particles. In other words, we may
put ${\cal M}_{\rm sup}=0$ for the ground level and then all
masses of quasi-particles and particles excited in the
tessellattice should be count off from that.

Another kind of vibrations may also be originated in the
tessellattice and this kind of vibrations is not typical for atoms
of the crystal lattice. In the presence of direct contact of the
superparticle surfaces one can presuppose that the collective
behavior of these surfaces should occur. According to what has
been said, Fig. 1 shows  states of vibrations of superparticles in
the space tessellattice.
\begin{figure}
\begin{center}
\includegraphics[scale=1.6]{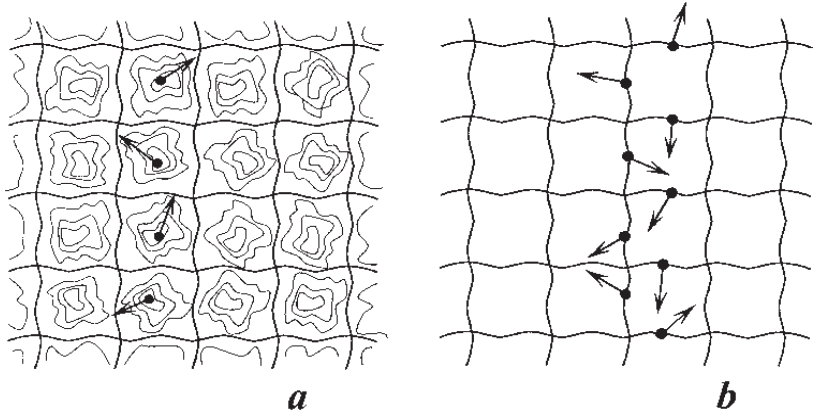}
\caption{\small Two kinds of superparticle vibrations in the
degenerate space: ($a$) vibrations of the centre-of-mass of
superparticles; ($b$) vibrations of the surface of
superparticles.} \label{Figure 1}
\end{center}
\end{figure}
In Fig. 1 {\kern 1pt}$a$ the center of gravity of each elementary
cell vibrates at the equilibrium position. In Fig. {\kern 1pt}
1$b$ the cellular structure is considered as a network and each
elementary link of the network vibrates at its equilibrium
position.

It is natural to assume that the reduction of the superparticle
volume from ${\cal V}_0$ in the free state to ${\cal V}_{\rm sup}$
in the degenerate state of the space (i.e. in the tessellattice)
takes place along with the conservation of the general area of the
superparticle's surface. Let ${\cal R}_0$ and ${\cal R}_{\rm sup}$
be typical radii of the same superparticle in the free state and
in the tessellattice, respectively. Then one can write the
relation
\begin{equation}
4\pi {\cal R}^2_0 =4\pi {\cal R}^2_{\rm sup} + \Delta S
\label {1}
\end{equation}
where the defect of surface $\Delta S$ of the superparticle can
probably be found in the state of a collective motion of fractal
elements. The above can schematically be shown in Fig. 2. Here the
defect of surface $\Delta S$ consists of the whole set of fractal
elements $\sigma_n$ (or amplitudes of the surface vibrations) that
cross inside and outside the surface formed by the radius ${\cal
R}_{\rm sup}$, such that $\Delta
S=\sum^{{\mathfrak{N}}}_{n=1}\sigma_n$. It is precisely these
surface vibrations that can be associated with fluctuations of the
electromagnetic field and the electric charge. Indeed the
orientation of the surface amplitudes $\sigma_n$ inside of the
superparticle (i.e., inside the sphere limited by the radius
${\cal R}_{\rm sup}$) we interpreted [29,30] as a quantum of
fractal deformations collapsed into one cell (the negative charge)
and the surface amplitudes $\sigma_n$ oriented outside the surface
of the superparticle is a quantum of fractal deformations, which
forms the positive charge. Thus we may conjecture that each
amplitude $\sigma_n$ oriented inside a superparticle describes the
negative element of the electric field and $\sigma_n$ oriented
outside of a superparticle depicts the positive element of the
electric field. Evidently, in the average the degenerate space
must be electro-neutral.
\begin{figure}
\begin{center}
\includegraphics[scale=3]{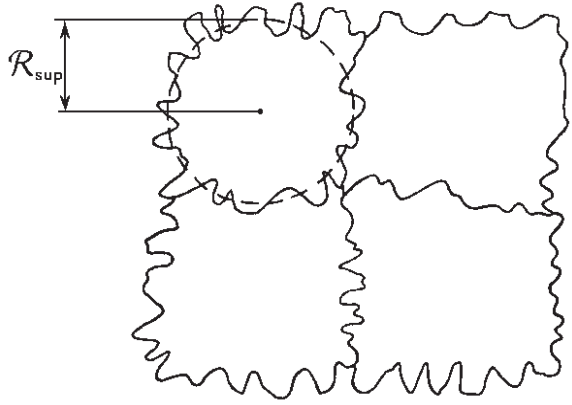}
\caption{\small Schematic presentation of collective vibrations of
the surfaces of superparticles in the degenerate space (in other
words, these are homeomorphic transformations of space elementary
cells).} \label{Figure 2}
\end{center}
\end{figure}
Therefore the sum of the amplitudes oriented towards the inside of
the superparticle surface has to be equal to the sum of the
amplitudes oriented towards the outside, $\sum_n \sigma_n^{(\rm
in)}=\sum_n \sigma_n^{(\rm out)}$.

Thus we arrive at the following statement, or axiom:

\medskip
\noindent \underline{\textbf{Axiom.}} The total surface of a
superparticle, whatever transformations of the superparticle may
occur, should be invariant.

\section*{\small 3. GEOMETRY OF THE CHARGE}

\hspace*{\parindent} The surface defect $\Delta S$ of a particle
may consist of exclusively unidirectional amplitudes
$\sigma_n^{(\rm out)}$ or $\sigma_n^{(\rm in)}$. That is why the
transformation of a superparticle to the positive charge can be
presented as follows
\begin{equation}
   4\pi{\cal R}^2_0 = 4\pi R^2_{\rm part} + \sum_{n=1}^{{\mathfrak{N}}}
    \sigma_n^{(\rm out)};
\label{2}
\end{equation}

\noindent  a similar relationship is true for the negative charge
(i.e., when all amplitudes $\sigma_n$ are $\sigma_n^{(\rm in)}$).
In view of the fact that the elementary charge is characterized by
the central symmetry one can suggest that all the amplitudes
$\sigma_n$ in (2) are identical in value. By setting
$\sigma_n=\sigma$ (here $\sigma \equiv\sigma^{(\rm out)}$ or
$\sigma \equiv \sigma^{(\rm in)}$) one rewrites (2) in the form
\begin{equation}
4\pi {\cal R}^2_0 =4\pi R^2_{\rm part} + {{\mathfrak{N}}} {\kern
1pt} \sigma. \label{3}
\end{equation}

\begin{figure}
\begin{center}
\includegraphics[scale=2.8]{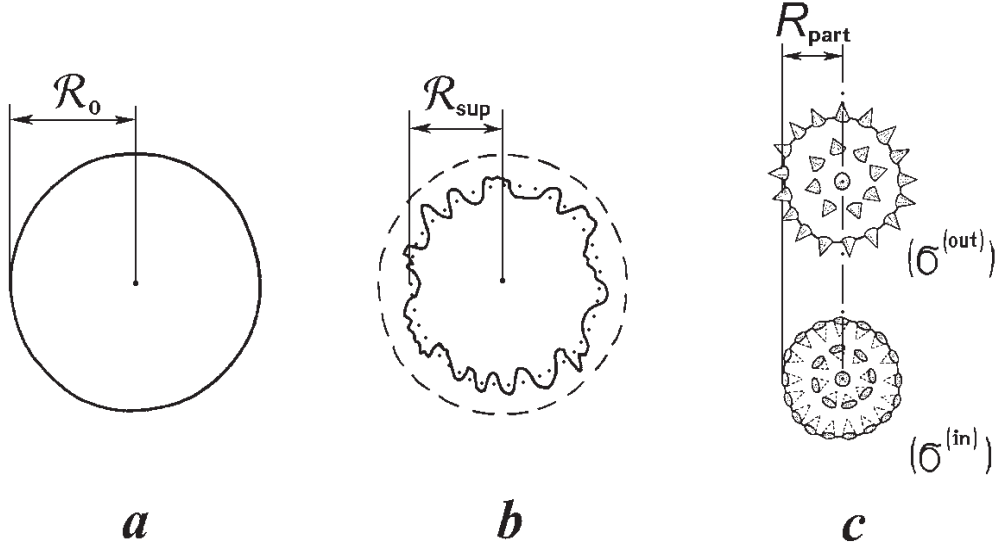}
\caption{\small Transitions of the superparticle from the free
state ($a$) to the degenerate state in the space tessellattice
($b$) and then to the state of the charged particle, either
positive or negative ($c$).} \label{Figure 3}
\end{center}
\end{figure}

\noindent
In particular, formula (3) should be justified for
leptons ($e, \ \mu, \ \tau$): with gradual increase of the mass
$M_{{\kern 1pt} e} < M_{{\kern 1pt}\mu} < M_{{\kern 1pt}\tau}$,
the radius must decrease ($R_{{\kern 1pt} e}>R_{{\kern 1pt}
\mu}>R_{{\kern 1pt} \tau}$) and the surface amplitude must
increase ($\sigma_e < \sigma_{\mu} < \sigma_{\tau}$). Transitions
of the superparticle from the free state to the degenerate state
in the tessellattice and then to the state of a particle are
schematically shown in Fig. 3.

Let us study the possibility of representing the surface defect
$\Delta S$ in the form shown in Fig. 3{\kern 2pt}{\it c}, which
can be called the "chestnut" model. In other words, the chestnut
model means that the particle is presented by a sphere with
needles that stick out (or inside) of the sphere surface.

Let the shape of the surface needle be defined by the figure of
rotation of the function
\begin{equation}
y = h \cdot \bigl( \frac {x}{r} - 1 \bigr)^2 \label{4}
\end{equation}
that revolves around the $OY$ axis, see Fig. 4 (another shape of
the needle is studied in Appendix). Here $h$ is the height of the
needle and $r$ is the radius of the base of the needle. Then the
area of surface of the needle is
\begin{eqnarray}
   \sigma_{\rm needle}=2\pi\int\limits^h_0 x(y) \ d l=
    2\pi \int\limits^h_0 r \Bigl( 1+ \sqrt{1+\frac {y}{h}} \ \Bigr)
    \sqrt{1+\frac {r^2}{4hy}} \ \  dy        \nonumber       \\
 =\frac {\pi r^4}{6h^2} \Bigl[ \Bigl( 1+{4h^2\over r^2}\Bigl)^{3/2}
   - 1 \Bigr] \ \ \ \ \ \ \ \ \ \ \ \ \ \ \ \ \ \ \ \ \ \ \ \ \ \
 \ \ \ \ \ \ \ \ \ \       \nonumber      \\
+\frac {\pi r^3}{2h} \Bigl[ \frac {2h}{r} \sqrt{1 + \frac
{4h^2}{r^2}} + \ln \Bigl( \sqrt{1+\frac {4h^2}{r^2}} +
\frac{2h}{r}
   \Bigr) \Bigr].
\label{5}
\end{eqnarray}
\begin{figure}
\begin{center}
\includegraphics[scale=3]{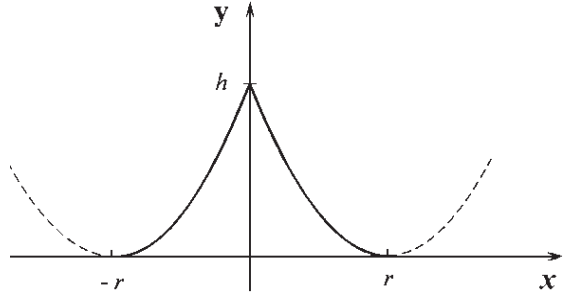}
\caption{\small Graphic display of the function (4).}
\label{Figure 4}
\end{center}
\end{figure}
Now we can write the total area of the particle surface:
\begin{equation}
S = 4\pi R^2_{\rm part}  - {{\mathfrak{N}}} {\kern 1pt} \sigma(r)
+ {{\mathfrak{N}}}{\kern 1pt} \sigma_{\rm needle}
 \label{6}
\end{equation}
where
\begin{equation}
\sigma(r) = 2\pi R^2_{\rm part} \Bigl( 1- \cos \frac {r}{R_{\rm
part}} \Bigr) \label{7}
\end{equation}
is the area of the needle base (i.e. circle) on the sphere with
the radius $R_{\rm part}$ [38]; it is obvious that the value of
${{\mathfrak{N}}}{\kern 1pt} \sigma (r)$ must be excluded from the
the surface defect $\Delta S$. Note that the part of the particle
surface, which is not overlapped by the needle bases (7), is
neglected. Since $r \lll R_{\rm part}$, one can reduce expression
(7) to $\sigma (r) \simeq \pi r^2$. Then in this case Eq. (6) due
to the axiom of the surface invariant is reduced to
\begin{equation}
S=4\pi R^{\kern 1pt 2}_{\rm part} + \Delta S, \label{8}
\end{equation}
\begin{eqnarray}
 \Delta S &=& {{\mathfrak{N}}} {\kern 1pt} \pi \Bigl\{
\frac {16 {\kern 1pt} r^4}{h^2} \Bigl[ \Bigl(1+\frac {4h^2}{r^2}
\Bigr)^{3/2} - 1 \Bigr]  \nonumber  \\ &+&\frac {r^2}{2h} \ln
\Bigl( \sqrt{1+\frac {4h^2}{r^2}} +\frac {2h}{r} \Bigr) +r^2 \sqrt
{1+ \frac {4h^2}{r^2}} - r^2  \Bigr\}. \label{9}
\end{eqnarray}
Let us look into the possibility of the  extreme of the surface
defect (9) that is a function of two parameters, $h$ and $r$. If
so, i.e., $\Delta S$ has the minimum at the defined variables
$h=h_{{\kern 1pt}0}$ and $r=r_0$, which are the solutions of the
equations
\begin{equation}
\Delta S^{{\kern 1pt}\prime}_h = 0; \label{10}
\end{equation}
\begin{equation}
     \Delta S^{{\kern 1pt}\prime}_r = 0,
\label{11}
\end{equation}
then the needle-shaped surface of the particle (Fig. 3,{\kern
3pt}{\it c}) will be the most stable among all the other possible
shapes. Necessary conditions for the existence of the minimum are
the inequalities (see, e.g. Refs. [39,40]):
\begin{equation}
{\rm Det} = \Delta S^{{\kern 1pt}\prime \prime}_{hh} (h_{{\kern
1pt}0}, r_0) {\kern 1pt} \Delta S^{{\kern 1pt} \prime \prime}_{rr}
(h_{{\kern 1pt}0}, r_0) - \Delta S^{{\kern 1pt} \prime
\prime}_{hr} (h_{{\kern 1pt}0}, r_0) {\kern 1pt} \Delta S^{{\kern
1pt} \prime \prime}_{rh} (h_{{\kern 1pt}0}, r_0)
> 0; \label{12}
\end{equation}
\begin{equation}
\Delta S^{{\kern 1pt} \prime \prime}_{hh} (h_{{\kern 1pt}0}, r_0)>
0, \ \ \ \ \ \Delta S^{{\kern 1pt} \prime \prime}_{rr} (h_{{\kern
1pt}0}, r_0)
>0. \label{13}
\end{equation}

     Eqs. (10) and (11) are reduced to, respectively,
\begin{equation}
\frac 1{\varkappa} \ln \Bigl( \sqrt {1+\varkappa^{{\kern 1pt} 2}}
+ \varkappa \Bigr) = \frac 13 \Bigl\{ 2+\sqrt{1+ \varkappa^{{\kern
1pt} 2}} -\frac 83 \frac 1{\varkappa^{{\kern 1pt} 2}} \Bigl[\Bigl(
1+ \varkappa^{{\kern 1pt} 2} \Bigr)^{3/2} -1 \Bigr] \Bigr\};
\label{14}
\end{equation}
\begin{equation}
\frac 1{\varkappa} \ln \Bigl( \sqrt{ 1+\varkappa^{{\kern 1pt} 2} }
+ \varkappa \Bigr) = 3 \sqrt {1+ \varkappa} - \frac 43 \frac
1{\varkappa^{{\kern 1pt} 2}} \Bigl[ \Bigl( 1 +\varkappa^2
\Bigr)^{3/2} - 1 \Bigr] \quad \quad \quad \label{15}
\end{equation}
where the denotation
\begin{equation}
\varkappa =2h/r \label{16}
\end{equation}
is introduced. Compatibility of Eqs. (14) and (15) is provided by
equating their right-hand sides. It gives the following equation
for $\varkappa$:
\begin{equation}
6 \varkappa^{{\kern 1pt} 2} + 5 = \bigl( \varkappa^{{\kern 1pt} 2}
- 5 \bigr) \sqrt {1 + \varkappa^{{\kern 1pt} 2} }. \label{17}
\end{equation}
The solution of Eq. (17) is
\begin{equation}
\varkappa_{{\kern 1pt} 0} \simeq 0.8889025353. \label{18}
\end{equation}
One finds the second derivatives of $\Delta S$:
\begin{eqnarray}
\Delta S^{ {\kern 1pt} \prime \prime}_{hh}= {\mathfrak{N}} {\kern
1pt} \pi \Bigl\{16 \frac 1{\varkappa^{{\kern 1pt} 4}} \Bigl[
\Bigl( 1+ \varkappa^{{\kern 1pt} 2} \Bigr)^{3/2} - 1 \Bigr]  \ \ \
\ \ \ \ \ \ \ \ \ \ \ \ \ \ \ \ \ \ \ \ \ \ \    \nonumber
\\  + 8 \frac 1{\varkappa^{{\kern 1pt} 3}} \ln \Bigl(
\sqrt{1 + \varkappa^{{\kern 1pt} 2}}+ \varkappa \Bigr) - 16 \frac
{2+ \varkappa^{{\kern 1pt} 2}}{\varkappa^{{\kern 1pt} 2}} \sqrt{1
+ \varkappa^{{\kern 1pt} 2}} \Bigr\}; \label{19}
\end{eqnarray}
\begin{eqnarray}
\Delta S^{{\kern 1pt} \prime \prime}_{rr}= {{\mathfrak{N}}} {\kern
1pt} \pi \Bigl\{8 \frac 1{\varkappa^{{\kern 1pt} 2}} \Bigl[ \Bigl(
1+ \varkappa^{{\kern 1pt} 2} \Bigr)^{3/2} -1 \Bigr]
 \ \ \ \ \ \ \ \ \ \ \ \ \ \ \ \ \ \  \ \  \  \ \ \ \ \ \ \
\nonumber   \\
+ 6 \frac 1 {\varkappa} \ln \Bigl( \sqrt{1 + \varkappa^{{\kern
1pt} 2}}+ \varkappa \Bigr) - 4 \frac {3+ 2 \varkappa^{{\kern 1pt}
2}}{\sqrt{1 + \varkappa^{{\kern 1pt} 2}}} - 2 \Bigr\}; \label{20}
\end{eqnarray}
\begin{eqnarray}
\Delta S^{{\kern 1pt} \prime \prime}_{hr}= {\mathfrak{N}} {\kern
1pt} \pi \Bigl\{-\frac {32}{3} \frac 1{\varkappa^{{\kern 1pt} 3}}
\Bigl[ \Bigl( 1+ \varkappa^{{\kern 1pt} 2} \Bigr)^{3/2} -1 \Bigr]
\ \ \ \ \ \ \ \ \ \ \ \ \ \ \ \ \ \ \ \ \  \  \ \ \
\nonumber
\\ - 6 \frac 1{\varkappa^{{\kern 1pt} 2 }} \ln \Bigl( \sqrt{1 +
\varkappa^{{\kern 1pt} 2}}+ \varkappa \Bigr) + 2 \frac {11+ 7
\varkappa^{{\kern 1pt} 2}}{\varkappa \sqrt{1 + \varkappa^{{\kern
1pt} 2}}} \Bigr\}; \label{21}
\end{eqnarray}
\begin{eqnarray}
\Delta S^{{\kern 1pt} \prime \prime}_{rh}= {{\mathfrak{N}}} {\kern
1pt} \pi \Bigl\{ -\frac {32}{3} \frac 1{\varkappa^3} \Bigl[ \Bigl(
1+ \varkappa^{{\kern 1pt} 2} \Bigr)^{3/2} -1 \Bigr]  \ \ \ \ \ \ \
\ \ \ \ \ \ \ \ \ \ \ \ \ \ \ \ \ \            \nonumber    \\ - 6
\frac1{\varkappa^{{\kern 1pt} 2} \ln} \Bigl( \sqrt{1 +
\varkappa^{{\kern 1pt} 2}}+ \varkappa \Bigr) + 2 \frac {11+ 10
\varkappa^{{\kern 1pt} 2}}{\varkappa \sqrt{1 + \varkappa^{{\kern
1pt} 2}}} \Bigr\}. \label{22}
\end{eqnarray}
By substituting the four derivatives (19)--(22) into inequalities
(12) and (13) we obtain at $\varkappa =\varkappa_{{\kern 1pt} 0}$
(see (18)), respectively:
\begin{eqnarray}
&{{\rm Det}|}_{\varkappa _{{\kern 1pt} 0}}&=({{\mathfrak{N}}}
{\kern 1pt} \pi)^2 \bigl\{ 2.645513 \times 3.835575 -
(-0.531893)\times 4.518111 \bigr\} \cr &&=({\mathfrak{N}}\pi)^2
\times 12.550216 >0; \label{23}
\end{eqnarray}
\begin{equation}
\Delta S^{\prime \prime}_{hh}(\varkappa_{{\kern 1pt} 0})=
{{\mathfrak{N}}} {\kern 1pt} \pi \times 2.645513 > 0, \ \ \ \ \ \
\Delta S^{\prime \prime}_{rr}(\varkappa_{{\kern 1pt}
0})={{\mathfrak{N}}} {\kern 1pt}  \pi \times 3.835575 > 0.
\label{24}
\end{equation}
As is seen from expressions (23) and (24), the necessary
conditions for the existence of the minimum of $\Delta S$ are
satisfied. Thus the function $\Delta S (h,r)$ (9) really has the
minimum at $h=h_{{\kern 1pt} 0}, \ r=r_{{\kern 1pt} 0}$.
Substituting respective values of  $\varkappa_{{\kern 1pt} 0}$ and
$h_{{\kern 1pt} 0}$ from (18) and (16) into expression (9) we can
rewrite expression (8) in the form
\begin{equation}
S=4\pi R^2_{\rm part} + {{\mathfrak{N}}} {\kern 1pt} \pi {\kern
1pt} r_0^2 \times 2.4157446 \label{25}
\end{equation}
where $r_0$ corresponds to the optimal radius of the needle base.
As evident from expression (25), the variation of $R_{\rm part}$
leads to the respective variation of $r_0$.

Let us derive the needle volume $V_{\rm needle}$. It can easy be
done in view of the rotation of the curve (4) around the $OY$
axis:
\begin{eqnarray}
V_{\rm needle}= \pi \int\limits^h_0 x^2(y) {\kern 1pt} d y = \pi
\int\limits^h_0 \bigl\{ r (1+\sqrt {y / h} \ ) \bigr\}^2 d y
\nonumber    \\
 = \frac {11}{12} {\kern
1pt} \pi r^2 h. \ \ \ \ \ \ \ \ \ \ \ \ \ \ \ \ \ \ \
  \label{26}
\end{eqnarray}
At point $(h_{{\kern 1pt} 0}, r_{{\kern 1pt} 0})$:
\begin{equation}
V ^{(0)}_{\rm needle}(h_{{\kern 1pt} 0}, r_{{\kern 1pt} 0})=
0.81482729 \times \pi r^3_0. \label{27}
\end{equation}
Clearly, the total volume of needles should satisfy the inequality
\begin{equation}
{{\mathfrak{N}}} V^{(0)}_{\rm needle} \ll \frac {4\pi}{3} R^3_{\rm
part}. \label{28}
\end{equation}
From here one obtains the restrictions to $r_0$:
\begin{equation}
r_0^3 \ll R^3_{\rm part}/{{\mathfrak{N}}}. \label{29}
\end{equation}
Inequality (29) with respect to the value $r_0$ is the distinctive
adiabatic condition: the two states of surface polarization (Fig.
3{\kern 1pt}{\it c}) of the particle can be considered as
practically stable only if $r_0$ is as small as the condition (29)
requires.

It follows from (29) that the surface defect  $\Delta S$ (25) that
generates the electric charge in a particle is significantly
smaller as compared with the ground surface $4\pi R^2_ {\rm
part}$. The surface $\Delta S$ covers  the volume $4\pi R^3_{\rm
part}/3$ responsible for gravitational properties of the particle.
However, on the other hand, the effective curvature  of the needle
is considerably greater than that of the ground surface of the
particle. That is why it is logical that the force on the side of
needles with the typical dimension $h_{{\kern 1pt} 0} \approx 0.45
{\kern 1pt} r_{{\kern 1pt} 0}$ acts upon surrounding
superparticles much strongly than that caused by the sphere with
the radius $R_{\rm part}$.

\section*{\small 5. ELECTRIC CHARGE AND ITS MOTION}

\hspace*{\parindent} It is obvious that each $n$th small needle on
the sphere surface can be regarded as the normal vector to the
particle surface. If we designate the normal dimensionless unit
vector as $\vec u$, the combination ${\vec u}{\kern 2pt} h_{\kern
1pt 0}/{\mathfrak{h}}$ can be interpreted as an elementary vector
of the electric field, i.e.
\begin{equation}
 {\vec {\cal E}}_{0n} = {\vec u}{\kern 2pt} \frac {h_{{\kern 1pt}
0}} {\mathfrak{h}}
 \label{30}
\end{equation}
where the normalized section ${\mathfrak{h}}$ plays the role of a
"quantum" of the length of the needle. The flux $U_n$ of the
vector ${\vec{\cal E}}_{0n}$ through the surface $\delta
s_{n0}\equiv \pi r^2_0$ of the $n$th needle base may be written in
the form of scalar product
\begin{equation}
U_n= {\vec {\cal E}}_{0n} \cdot \delta {\vec s}_{n0}.
 \label{31}
\end{equation}
Then the elementary electric charge $\tilde e$ can be determined
as the total sum of all fluxes $U_n$ through the particle surface:
\begin{equation}
\tilde e = \sum_{n=1}^{{\mathfrak{N}}} U_n ={{{\mathfrak{N}}}}
\Bigl(
       {\vec {\cal E}}_{0n} \cdot \delta {\vec s}_{n0} \Bigr).
\label{32}
\end{equation}
Note that a constancy of the charge $\tilde e$, for instance, for
the lepton series, should be provided by increasing the radius
$r_0$ when $R_{\rm part}$ decreases: $r_{0e}< r_{0 \mu}< r_{0
\tau}$ when $R_e> R_{\mu}> R_{\tau}$. On the background of the
noise the stable polarization of the "chestnut" (i.e., a quantum
of fractal deformations) in the state of rest induces the same
polarization  on surrounding superparticles, such that two
adjacent superparticles exhibit opposite forms: one in the sense
of convexity and the other in the sense of concavity (Fig. 5).
\begin{figure}
\begin{center}
\includegraphics[scale=1.5]{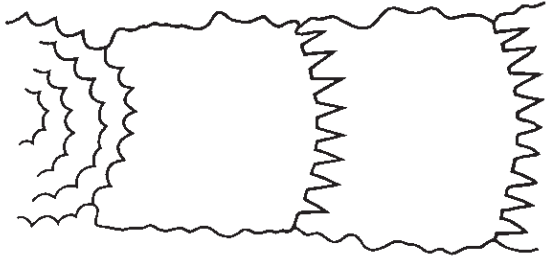}
\caption{\small Superparticles polarized by the positive charged
particle-"chestnut". The positive polarization induced in the
superparticles is directed outside the superparticles (needles
out); the negative polarization is directed inside the
superparticles (needles in).} \label{Figure 5}
\end{center}
\end{figure}
The neighboring superparticles transmit the particle  polarization
to the adjacent superparticles up to the boundary of the
crystallite whose size is defined by the Compton wavelength of the
particle $\lambda_{\rm {\kern 1pt} C} = h/Mc$ [21] and hence the
field ${\vec {\cal E}}_{0n}$ shown in Fig. 5 extends for a
distance of the radius $\lambda_{\rm {\kern 1pt} C}$ from the
particle.

     Let us assume that the height of the needle can vary (for example,
it can oscillate between magnitudes $h_{{\kern 1pt} 0}$ and
$h_{{\kern 1pt} 0} -\delta h_{{\kern 1pt} 0}$). In this case each
of the needle states  has its own surface stretched on the same
base. Such kind of the needle motion is potential and hence all
states of the needle surface can be described by a scalar function
$\Phi_n (h)$. Then the field vector ${\vec{\cal E}}_n$ can be
associated with the scalar function $\Phi_n(h)\propto h$, i.e.
\begin{equation}
{\vec {\cal E}}_n(h) = - \nabla_h \Phi_n(h)
 \label{33}
\end{equation}
(so ${\vec{\cal E}}_n$ is a co-vector).

    On the other hand, the spike of each $n$th needle is able to
deviate from its equilibrium position, i.e., the bending of the
needle from its axis of symmetry must not be ruled out. It is
obvious that the value of displacement decreases from the spike to
the base of the needle, which is fixed. Therefore this kind of
motion can be related to a vector field rather than to a vector
that is true only for the displacement of a separate point. Let us
designate this field as ${\vec{\cal A}}_n$.

     Let us treat the motion of the "chestnut" along its spatial
trajectory $l$. The interaction of a moving particle with
superparticles results in a very specific dynamics of the particle
along its path [19-23]. In those works we have studied features
associated with the translation motion and the intrinsic motion
(i.e., an asymmetric pulsation of the center-of-mass of a
particle). There was no need to introduce the rotational degree of
freedom in our previous papers. However, there is a good reason to
do it now: It is precisely this motion of the charge that the
rotational electric and magnetic fields generate.

   The initial condition is the significant characteristic of the
motion. In preceding papers [19-23], an initial spatial velocity
${\vec v}_0$ of the particle was regarded as a main property. The
value of $v_0$ described both the translation movement and the
intensity of the proper pulsation of the particle. Besides we
assumed that the velocity $v_0$ was an essential characteristic of
the motion of elements of the particle surface as well. Aside from
the value of $v_0$, which has to pose a dynamics for all the three
kinds of degrees of freedom of the particle, the intrinsic motion
and the surface motion in addition should have some initial
conditions. The direction of the displacement of the
center-of-mass of the particle in relation to the vector ${\vec
v}_0$ should ensure the condition for the intrinsic motion (spin),
i.e., along or against  $\vec v_0$. In the case of the surface
motion, we may assume that the direction of the field ${\cal A}_n$
of each $n$th needle is transversal to the vector ${\vec v}_0$.
Besides, the vector ${\vec {\cal A}}_n$ can be oriented either to
the right or to the left relative to ${\vec v}_0$ (compare with
the photon, a carrier of the electrodynamic field, which is
characterized by the transversal polarization, either right or
left in relation to its path).

Let us extend the major features of the dynamics of a chargeless
particle [19-23] to the charged one. That is, one can suppose that
with the motion of the "chestnut", respective magnitudes of the
scalar and vector fields, $\Phi_n$ and ${\vec {\cal A}}_n$, for
each  $n$th needle oscillate with the period of collisions $T$ in
time and with the period $\lambda$ in space (here $\lambda
=v_0T$). Thus with each $i$th collision of the particle with an
ingoing superparticle, the former emits an inerton. With regard
for the availability of the charge state on the superparticle
surface, this inerton also carries over a bit of the
electromagnetic field. [Recall that in solid state physics the
mixture of two different quasi-particles in one entity is of
frequent occurrence; in particular, composite quasi-particles can
be attributed to the polariton (the mixture of light and polarized
optical phonons), the polaritonic exciton (the mixture of the
polariton and exciton), the excitonic polaron (the mixture of the
polaron and exciton), etc.] The pattern of the motion of a charged
particle shrouded in its inerton-photon cloud is schematically
illustrated in Fig. 6. The particle emits inerton-photon
quasi-particles that move from superparticle to superparticle by
the relay mechanism. Within the period $T_i$ of the collision
between the particle and the $i$th emitted inerton-photon, a part
of the velocity vector ${\vec v}_{0i}$ is transmitted from the
particle to the quasi-particle and remains constant in the latter.
\begin{figure}
\begin{center}
\includegraphics[scale=1.35]{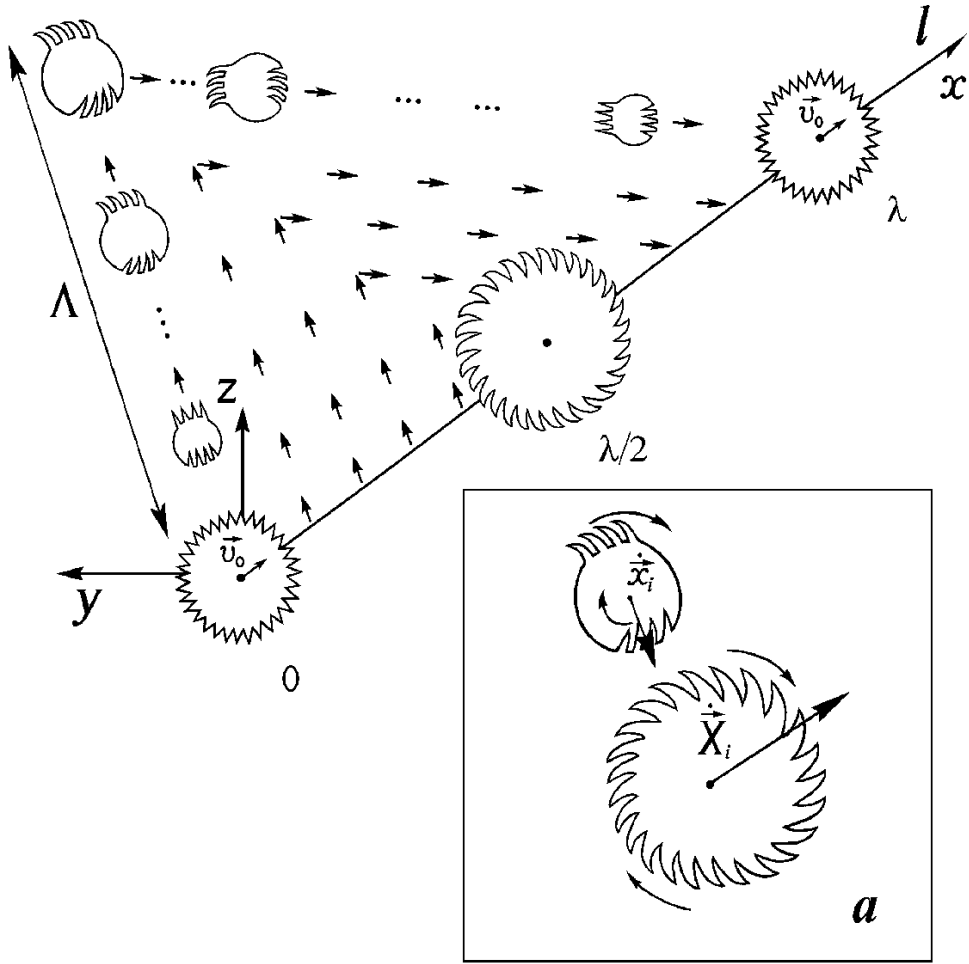}
\caption{\small Diagram of the motion of the positive charged
particle. The particle is accompanied by electromagnetic polarized
inertons, or inerton-photons, or simply photons (it is obvious
that these particle's polarized inertons correspond to so-called
"virtual photons" of quantum electrodynamics). $(a)$ the moment of
absorption of the $i$th inerton-photon by the particle.}
 \label{Figure 6}
\end{center}
\end{figure}
However, phase states of the $i$th quasi-particle at the moments
$t_i=T_i \aleph$ where $ \aleph =0$, 1, 2, ... and $t _i=(2 \aleph
+ 1)T_i/2$ are absolutely different [24] (here $\aleph$ designates
the order number of the de Broglie section $\lambda$ in a whole
particle path $l$; this is because the length of the particle path
$l$ is subdivided into a set of sections $\lambda$). At the moment
$t_i=0$, the quasi-particle possesses the inert mass and the
electric polarization and the motion of the center-of-mass and the
surface polarization (i.e., the rotation of the electric field)
begins from that. At the moment $t_i=T_i/2$, the distance between
the particle and the quasi-particle is equal to $\Lambda_i$ and
the quasi-particle does not possess any mass in this point (the
mass is transformed here to a local deformation of the
tessellattice as a whole called the 'rugosity' in Ref. [24]). The
deformation is anisotropic; in this position the quasi-particle's
spin is found in the explicit form [21,24]. Moreover, the
anisotropic state is specified by the axial polarization that can
be considered as a bit of the magnetic field.

The same happens to the particle:  at the moment $t=0$, it is
characterized by the initial velocity of translational motion
$v_0$, by  the inert mass $M_0/\sqrt{1-v_0^2/c^2}$, by the normal
polarization, or fractality of the surface (the electric charge),
by the initial velocity $v_0$ of the intrinsic motion (the motion
of center-of-mass) and by the initial velocity $v_0$ of the motion
of the surface polarization. At the moment $t=T/2$, the particle
is found at rest: its velocity of the translational motion is
equal to zero, the mass equals 0 as well, the center-of-mass is
displaced from the equilibrium position (there is the spin) and
the particle surface is specified by the axial polarization (i.e.,
one can see the explicit magnetic charge -- the monopole, Fig. 6).

Let us analyze the motion of the scalar $\Phi_n$ and vector ${\cal
A}_n$ fields mathematically. Dimensionality of these fields
conforms to length. Therefore, the corresponding velocities are
${\dot \Phi}_n$ and ${\dot {\cal A}}_n$. For the sake of
simplicity we will describe the inerton-photon cloud as a whole
quasi-particle. Let $\phi_n$ and ${\vec \alpha}_n$ be the scalar
and vector fields, respectively, of the $n$th effective needle of
this quasi-particle; hence the velocities of the given fields are
${\dot \phi}_n$ and ${\dot {\vec \alpha}}_n$.  Let us write the
Lagrangian density for the motion of the $n$th particle's needle
and the $n$th cloud's needle taking into account their mutual
interaction:
\begin{eqnarray}
{\cal L}_n&=&C \Bigl\{{1\over 2}{\kern 1pt}{\dot \Phi}^2_n + \frac
12 {\kern 1pt} {\dot{\vec {\cal A}}}^2_n + {1\over 2}{\kern
1pt}{\dot \phi}^2_n +\frac 12 {\kern 1pt}{\dot {\vec
\alpha}}^2_{{\kern 1pt} n} \nonumber
\\ &-& v_0 \Bigl({\dot \Phi}_n \nabla {\vec \alpha}_{{\kern 1pt} n}
+{\dot \phi}_n \nabla {\vec {\cal A}}_n \Bigr) - v_0^2 \Bigl(
\nabla \times {\vec {\cal A}}_n \Bigr) \Bigl( \nabla \times {\vec
{\alpha}}_{{\kern 1pt}n} \Bigr) \ \Bigr\} \label{34}
\end{eqnarray}
($C$ is a constant, its dimensionality in SI is kg/m$^3$). Here
quadratic forms in the first line correspond to the kinetic energy
of the fields $\Phi_n$ and ${\vec {\cal A}}_n$ of the $n$th
particle's needle and the kinetic energy of the fields $\phi$ and
${\vec \alpha}_n$ of the $n$th effective needle of the cloud. The
next to the last and the last terms describe the interaction
between the fields of the particle and the cloud. In expression
(34), the fields are differentiated by the time $t$ that is set as
the proper time of the particle, i.e., $t=l/v_0$ is a natural
parameter, because $l$ is the length of the particle path [20]. It
is important to emphasize that owing to the introduction to the
Lagrangian density (34) such operators as the divergence and the
curl the derivatives $\dot{\Phi}_n$, \ $\dot{\phi}_{{\kern 1pt}
n}$ and ${\dot{\vec{\cal A}}}_{{\kern 1pt} n}$, \
$\dot{\vec{\alpha}}_{{\kern 1pt} n}$ should be treated as the
partial derivatives: $\dot{\Phi}_n =
\partial {\Phi_n}/\partial {\kern 1pt} t$, \ $\dot{\phi}_{{\kern 1pt}
n} =\partial {\phi_{{\kern 1pt} n}}/\partial {\kern 1pt} t$ \ and
\ ${\dot{\vec{\cal A}}}_{{\kern 1pt} n} =\partial{{\vec{\cal
A}}}_{{\kern 1pt} n}/\partial {\kern 1pt} t$, \
$\dot{\vec{\alpha}}_{{\kern 1pt} n}=\partial{\vec{\alpha}}_{{\kern
1pt} n}/\partial {\kern 1pt} t$. This means that these time
derivatives describe changes of the corresponding fields in those
points $\vec r$ of the particle trajectory $l$ in which
divergences $\nabla{\vec{A}}_n$ and $\nabla{\vec{\alpha}}_n$ and
vorticities $\nabla \times{\vec{A}}_n$ and $\nabla
\times{\vec{\alpha}}_n$ are taken. The operator $\nabla$ in
expression (34) is constructed on the spatial coordinate $\vec r$
that is defined as $\vec r = \vec \ell + \overrightarrow {(R_{\rm
part} + h)}$, where $\vec\ell$ is the radius vector of equilibrium
position of the center-of-mass of the particle on the path $l$;
$\overrightarrow {(R_{\rm part} + h)}$ is the vector eliminating
from the equilibrium position of the center-of-mass of the
particle to the spike of the $n$th needle ($\vec\ell$ and $h$
change along the particle trajectory, but $R_{\rm part}$ is
considered as a constant, though it is reduced along the
trajectory owing to the so-called relativistic effect [20]. So
$\vec r$ is the radius vector of the location of the spike of the
$n$th needle. It is obvious that $\vec r$ describes the sharp
location of the $n$th  needle of the cloud.

We can put the following relations that describe the local
electro-neutrality:
\begin{equation}
\Phi_n + \phi_{{\kern 1pt} n} = 0, \ \ \ \ \ \ \ \ {\vec{\cal
A}}_n + {\vec \alpha}_{n} =0. \label{35}
\end{equation}
Then the Lagrangian density (34) may be presented as
\begin{equation}
{\cal L}_n = {\cal L}_n^{\rm part} + {\cal L}_n^{\rm cloud};
\label{36}
\end{equation}
\begin{equation}
\ \ \  \ \ \ \ \ \ \ \ \ \ \ \ {\cal L}_n^{\rm part}=C \Bigl\{
{1\over 2}{\kern 1pt} {\dot \Phi}_n^{{\kern 1pt} 2} + \frac 12
{\kern 1pt} {\dot {\vec {\cal A}}}_n^{\ 2} + v_0{\dot \Phi}_n
\nabla {\vec {\cal A}} _n + \frac 12 {\kern 1pt} v_0^2 \bigl(
\nabla \times {\vec {\cal A}}_n \bigr)^2 \Bigr\};   \label{37}
\end{equation}
\begin{equation}
{\cal L}_n^{\rm cloud}=C \Bigl\{ {1\over 2}{\kern 1pt} {\dot
\phi}^{{\kern 1pt} 2}_n + \frac 12 {\kern 1pt} {\dot {\vec
\alpha}}^{{\kern 1pt} 2}_{{\kern 1pt} n} + v_0 {\dot \phi}_{{\kern
1pt} n} \nabla {\vec\alpha}_{{\kern 1pt} n} + \frac 12 {\kern 1pt}
v_0^2 \bigl( \nabla \times {\vec \alpha}_{{\kern 1pt} n} \bigr)^2
\Bigr\}. \quad \ \label{38}
\end{equation}
     Since the Lagrangian densities (37) and (38) have the same form,
one can treat the behavior of the potentials of the  particle's
needle only. The Lagrangian density (37) is a function of ${\dot
\Phi}_n$, $\nabla \Phi_n$ and ${\ \dot {\vec {\cal A}}}_n$,
$\nabla {\vec {\cal A}}_n$. In this case, the Euler-Lagrange
equations take the form (see, e.g. ter Haar [41]):
\begin{equation}
\frac {\partial}{\partial t} \frac {\partial {\cal L}}{\partial
{\dot Q}} -\frac {\delta {\cal L}}{\delta Q}= 0 \label{39}
\end{equation}
where the functional derivative
\begin{equation}
\frac {\delta {\cal L}}{\delta Q} = \frac {\partial {\cal
L}}{\partial Q} - \frac {\partial}{\partial x} \frac {\partial
{\cal L}} {\partial \bigl( \frac {\partial Q }{\partial x} \bigr)
} - \frac {\partial}{\partial y} \frac {\partial {\cal L}}
{\partial \bigl( \frac {\partial Q}{\partial y} \bigr) } - \frac
{\partial}{\partial z} \frac {\partial {\cal L}} {\partial \bigl(
\frac {\partial Q}{\partial z} \bigr)}.
 \label{40}
\end{equation}
For the Lagrangian density (37) Eq. (39) gives the following two
equations:

\begin{equation} {\ddot \Phi}_n + v_0  \nabla  {\dot
{\vec {\cal A }}}_{{\kern 1pt} n}  = 0;
 \label{41}
\end{equation}

\begin{equation}
{\ddot {\vec {\cal A}}}_n + v_0 {\kern 1pt} \nabla {\dot \Phi}_n +
v_0^2 {\kern 1pt} \nabla \times \bigl( \nabla \times {\vec {\cal
A}}_n \bigr) = 0. \label{42}
\end{equation}
From Eq. (41) we obtain
\begin{equation}
{\dot \Phi}_n = - v_0 \nabla {\vec {\cal A}}_n + {\rm const}.
\label{43}
\end{equation}
Substituting ${\dot \Phi}_n$ from Eq. (43) into Eq. (42) we get
\begin{equation}
{\ddot {\vec {\cal A}}}_n - v_0^2 {\kern 1pt} \nabla \bigl( \nabla
{\vec {\cal A}}_n \bigr) + v_0^2 {\kern 1pt} \nabla \times \bigl(
\nabla \times  {\vec {\cal A}}_n \bigr) = 0. \label{44}
\end{equation}
With the known formula
\begin{equation}
\nabla \bigl( \nabla {\kern 1pt} \vec b {\kern 1pt} \bigr) -
\nabla \times \bigl( \nabla \times \vec b {\kern 1pt} \bigr) =
\nabla ^{2} {\kern 1pt} \vec b \label{45}
\end{equation}
we can write instead of Eq. (44)
\begin{equation}
{\ddot {\vec {\cal A}}}_n - v_0^2 {\kern 1pt} \nabla^{{\kern 1pt}
2} {\vec {\cal A}}_n = 0. \label{46}
\end{equation}

Integrating Eq. (42) over $t$ we can solve the equation relative
to the variable ${\dot{\vec {\cal A}}}_n$. By substituting the
obtained expression for ${\dot{\vec {\cal A}}}_n$ into Eq. (41) we
come to equation
\begin{equation}
{\ddot \Phi}_n - v_0^2 \nabla^{{\kern 1pt}2} \Phi_n - v_0^3 \Bigl(
\nabla \times \bigl( \nabla \times \int {\vec {\cal A}}_n {\kern
1pt} d t \bigr) \Bigr) = 0.
 \label{47}
\end{equation}
Since the formula
\begin{equation}
\nabla \bigl(\nabla \times \vec b {\kern 2pt}\bigr) = 0
 \label{48}
\end{equation}
is valid for any vector field $\vec b$, the last term in Eq. (47)
equals zero and we arrive at equation
\begin{equation}
{\ddot \Phi}_n - v_0^2 {\kern 1pt} \nabla^{{\kern 1pt}2} \Phi_n =
0. \label{49}
\end{equation}

The same equations can be derived for the fields ${\vec \alpha}_n$
and $\phi_n$ from the Lagrangian density (38).

The total fields of the particle and the cloud are respectively:
\begin{equation}
\Phi = \sum_{n=1}^{{\mathfrak{N}}} \Phi_n, \ \ \ \ \ \ {\vec {\cal
A}}= \sum^{{\mathfrak{N}}}_{n=1}{\vec {\cal A}}_n;
 \label{50}
\end{equation}
\begin{equation}
\phi= \sum^{{\mathfrak{N}}}_{n=1}\phi_n, \ \ \ \ \ \ \ \vec \alpha
= \sum^{{\mathfrak{N}}}_{n=1}{\vec \alpha}_n.
 \label{51}
\end{equation}
Evidently there is a phase lag between the fields $\Phi$ and
${\vec{\cal A}}$ (analogously for $\phi$ and $\vec \alpha$). The
value of the lag is equal to $\pi/2$. As seen from the wave
equations (46)  and (47), the rate of change for each of the
fields is defined by the velocity  $v_0$ of the spatial motion of
the particle. The initial conditions should be the following. For
the scalar field: $\Phi(0)=\Phi_0$, i.e.,  at the moment $t=0$,
the charge state of the particle surface is entirely characterized
by the central symmetry (${\mathfrak{N}}$ elementary vectors stick
out of the surface, ${\vec {\cal E}}_{0n}=-\nabla {\Phi_n
\vert}_{h=h_0}$), at the moment $t=T/2$ (or the spatial half-
period $\lambda/2$), the field $\Phi$ reaches the minimum, rather
zero; then at the moment $t=T$ (or the spatial period $\lambda$),
$\Phi (T)= \Phi_0$ again and so on. For the vector field: ${\ \dot
{\vec {\cal A}}}(0)= {\vec \epsilon}_{\pm} {\kern 1pt} v_0$
[recall that the dimensionality of the fields $\vec {\cal A}$ and
$\vec {\alpha}$ is length] where ${\vec \epsilon}_{\pm} =\pm 1$ is
the polarization vector; this condition implies that at the moment
$t=0$ all the $\mathfrak{N}$ needles on the sphere take the same
tangential velocity (to the right for $\epsilon_{+}=1$ or to the
left for $\epsilon_{-}= -1$ relative to the particle path $l$) and
begin to move synchro. At the moment $t=T/2$, the tangential
velocity of the needles decreases to zero, i.e., at this moment
the deviation, or more exactly, the bending of the needles are
maximal and the charge state of the particle surface is entirely
characterized by the axial symmetry (Fig. 6). Then at $t=T$, one
has ${\dot {\vec {\cal A}}}(T) = {\vec \epsilon_{\pm}}{\kern 1pt}
v_0$ and so on.

\section*{\small 6. THE CHARGE IN ELECTROMAGNETIC FIELD}

\vspace{2mm} \hspace*{\parindent} Let us examine a flux of free
photons, elementary excitations of space,  which carry over scalar
and vector polarizations, i.e., the field potentials $\phi$ and
$\vec \alpha$, from superparticle to superparticle by means of the
relay mechanism. Clearly the motion of the potentials $\phi$ and
$\vec \alpha$ of a free photon along its trajectory is described
by Eqs. (49) and (46), respectively; besides, in this case the
velocity $v_0$ should be replaced by the speed of light $c$.
However, the frequency $\nu$ of the phase alteration of the
potentials $\phi$ and $\alpha$ is given by a particle or
particles, which irradiated these photons ($\nu=1/2T$ where $T$ is
the period of collisions between the particle and its own
inerton-photon cloud [19-21]). The Lagrangian density ${\cal
L}^{\rm cloud}$ of a free photon has the form of expression (38)
in which the following substitutions should be made: $\phi_n
\rightarrow \phi$, ${\vec \alpha}_n \rightarrow {\vec \alpha}$ and
$v_0 \rightarrow c$. So
\begin{equation}
{\cal L}_{\rm photon}= C \Bigl\{ \frac 12 {\kern 2pt} {\dot
\phi}^{{\kern 1pt}2} + \frac 12 {\kern 2pt} {\dot{\vec
\alpha}}^{\kern 0.5 pt 2} + c{\kern 1pt}{\dot \phi} {\kern 1pt}
\nabla {\vec \alpha} + \frac 12 {\kern 2pt} c^{{\kern 0.5pt} 2}
\bigl( \nabla \times {\vec \alpha} \bigr)^2 \Bigr\}
 \label{52}
\end{equation}

       The elementary charge $\tilde e$ is organized by the flux of
the gradient of the field $\Phi$ of the particle, see formulas
(32) and (33), therefore, the charge $\tilde e$ can fully interact
with external photons, i.e., with the potential $\phi$. Then the
term of the Lagrangian density, which has to describe the
interaction, can be taken in the form
\begin{equation}
- C {\kern 1pt} {\tilde e}  \times \phi {\kern 2pt} \frac {c^2}
{\cal W} \label{53}
\end{equation}
where the factor $c^2/{\cal W}$ is introduced to guarantee the
dimensionality required, i.e., the energy density, of the term
(recall that the dimensionality of the value $\tilde e$ is the
area, m$^2$, see (32)). The dimensionality of the parameter $\cal
W$ is the volume, m$^3$ and therefore it is an effective volume
occupied by the charge $\tilde e$ at its interaction with external
photons.

    Let us determine the charge jointly with its deformation coat,
or the space crystallite and consider the inerton-photon cloud as
the whole object. In that event the object will be characterized
by the constant velocity ${\vec v}_0$ in the space and then the
flux ${\tilde e}{\kern 1pt}{\vec v}_0$ should be treated as a
current. The current as a vector is capable to interact with an
external vector field $\vec \alpha$. So the density of the
interaction Lagrangian should be supplemented with one more term,
\begin{equation}
C {\kern 1pt} {\tilde e} {\kern 1pt} {\kern 1pt} {\vec v}_0 \cdot
{\vec \alpha} {\kern 1pt} \frac {c}{\cal W}.
 \label{54}
\end{equation}

Let us now pass on to values whose dimensionality is traditional
in physics. In the System International: the charge $e$ is
measured in C (Coulomb); the scalar potential $\varphi$ in
kg$\cdot$m$^2$/(C$\cdot$s$^2$); the vector potential $\vec A$
{\kern 1pt} in kg$\cdot$m/(C$\cdot$s); the electric constant
$\varepsilon_0$ in C$^{{\kern 1.5pt}
2}$$\cdot$s$^2$/(kg$\cdot$m$^3$). Having passed to these standard
physical values $\varphi$, $\vec A$ and $e$, we should introduce a
dimensional parameter $\eta$, such that
\begin{equation}
{\vec A}=\eta {\vec \alpha}, \ \ \ \ \ \  \varphi=\eta c \phi.
\label{55}
\end{equation}
Using parameters $\cal W$ and $\eta$ introduced above, the
electric charge $e$ and its density can be presented as follows:
$e={\tilde e}{\kern 1pt}\varepsilon_0 {\kern 1pt} c {\kern 1pt}
\eta$ and $\rho = e/{\cal W}$ respectively. Besides, the constant
$C$ entering the Lagrangian density (34), (36)--(38) and
(52)--(54) can be written as $C=\varepsilon_0 \eta^{{\kern 1pt}
2}$. Then in standard symbols the Lagrangian density of the
electromagnetic field that interacts with a charge particle takes
the form
\begin{equation}
{\cal L}= \frac {\varepsilon_0}{2{\kern 1pt}c^2} {\kern 2pt} {\dot
\varphi}^{{\kern 0.5pt}2} + \frac {\varepsilon_0}{2} {\dot {\vec
A}}^{{\kern 1pt} 2} + \varepsilon_0 {\kern 0.4pt} {\dot
\varphi}{\kern 0.3pt} \nabla {\vec A} + \frac {\varepsilon_0
{\kern 1pt} c^2}{2} {\kern 2pt} \bigl( \nabla \times {\vec A}
{\kern 1pt} \bigr)^2 - \rho {\kern 1pt} \varphi + \rho {\kern 1pt}
{\vec v}_0 {\vec A}.
 \label{56}
\end{equation}
Note that the standard Lagrangian of the electromagnetic field
does not contain ${\dot \varphi}$, because they do not know in
what way it can be introduced (see, e.g. ter Haar [42]).

Euler-Lagrange equations (39) based on the Lagrangian density (56)
culminate in the following equations for the scalar $\varphi$ and
vector $\vec A$ potentials of the electromagnetic field
\begin{equation}
{\ddot \varphi} + c^2 \nabla {\dot {\vec A}} + \varepsilon_0
{\kern 1pt} c^2 \rho =0;
 \label{57}
\end{equation}
\begin{equation}
{\ddot {\vec A}} + \nabla {\dot \varphi} + c^2 {\kern 1pt} \nabla
\times \bigl( \nabla \times {\vec A} {\kern 1pt} \bigr) - \rho
{\kern 1pt} {\vec v}_0 =0.
 \label{58}
\end{equation}
Eqs. (57) and (58) are reduced to the two well-known D'Alambert
equations, Tamm [43], for $\varphi$ and $\vec A$:
\begin{equation}
\nabla^{{\kern 1pt} 2} \varphi - \frac 1{c^2} {\kern 1pt} \frac
{\partial^{{\kern 1pt} 2} {\kern 1pt} \varphi}{\partial {\kern
1pt} t^{{\kern 0.3pt}2}}
 = -\frac {\rho}{\varepsilon_0};
  \label{59}
\end{equation}
\begin{equation}
\nabla^{{\kern 1pt} 2} {\vec A}
  -\frac 1{c^2} \frac {\partial ^{{\kern
1pt} 2}{\vec A}}{\partial{\kern 1pt} t^2} = -\frac {\rho {\kern
1.3pt} {\vec v}_0}{\varepsilon_0 c^2}.
 \label{60}
\end{equation}
Eqs. (59) and (60) are the consequence of the Maxwell equations if
the electric field $\vec E$ and the magnetic induction $\vec B$
are associated with the potentials $\varphi$ and $\vec A$ by the
formulas
\begin{equation}
\vec E= - \nabla \varphi - \frac {\partial \vec A }{\partial
{\kern 1pt}t}, \ \ \ \ \ \ \vec B = \nabla \times {\vec A}.
 \label{61}
\end{equation}

   The Lagrangian of a free photon can be obtained from expression
(56) if one drops the two last terms. The equations of motion of
the photon are
\begin{equation}
{\ddot \varphi} + c^2 {\kern 1pt} \nabla {\dot {\vec A}} =0;
 \label{62}
\end{equation}
\begin{equation}
{\ddot {\vec A}} + \nabla {\kern 1pt} {\dot \varphi} + c^2 {\kern
1pt} \nabla \times \bigl( \nabla \times {\vec A} {\kern 1pt}
\bigr) =0,
 \label{63}
\end{equation}
which are reduced to the two wave equations for the photon
potentials
\begin{equation}
\nabla^2 \varphi - \frac 1{c^2} \frac {\partial^{{\kern 1pt} 2}
\varphi}{\partial {\kern 1pt} t^2}
 = 0; \label{64}
\end{equation}
\begin{equation}
\nabla^2 {\vec A}
  -\frac 1{c^{{\kern 0.3pt}2}} \frac {\partial ^{{\kern 1pt} 2}{\vec A}}
{\partial {\kern 1pt} t^2} = 0. \label{65}
\end{equation}

At the same time the migration of the photon "core", which occurs
by the relay mechanism from cell to cell, obeys equation [35]
\begin{equation}
 \frac{d}{d {\kern 1pt} t}{\kern 2pt}{\vec l}= \frac {\vec l}{l}
{\kern 2pt} c     \label{66}
\end{equation}
where $\vec l$ is the radius vector of the photon location and the
equation is held for the proper time $t > t_0$ where $t_0$ is the
"inoculating" time, i.e. the photon lifetime in one cell. Thus
equations (64), (65) and (66) completely describe the behavior of
the photon.

Note that the model leads to the Lorentz calibration
automatically, because the calibration immediately follows from
Eq. (62):
\begin{equation}
\nabla {\vec A}=-\frac{1}{c^2}\frac{\partial\varphi}{\partial t}.
 \label{67}
\end{equation}

The Maxwell equations act in the area of classical physics, i.e.,
when wave properties of a charged particle can be neglected.
Evidently, this statement is the direct consequence of the
Lagrangian density written in the form (56), which takes into
account the interaction of the external field with the whole
\{particle + deformation coat + inerton-photon cloud\} complex.
The complex moves as a classical object and its intrinsic dynamics
does not appear beyond the cloud boundary. The cloud is restricted
along the particle trajectory by the spatial period (i.e., de
Broglie wavelength) $\lambda$ and in transversal directions to the
trajectory by the cloud amplitude  $\Lambda \sim \lambda {\kern
1pt} c/v_0$. Thus the two particle parameters $\lambda$ and
$\Lambda$ limit the application of classical electrodynamics. Yet
the wavelength $\lambda_{{\kern 1pt} \rm photon}$ of external
photons (a section of a photon path or, in other words, the
spatial period of the phase change of the potentials $\varphi$ and
$\vec A$) [35] and the time period $T_{\rm photon}=\lambda_{{\kern
1pt} \rm photon}/c$  \ should satisfy the inequalities
\begin{equation}
\lambda_{{\kern 0.3pt}\rm photon} \gg \Lambda, \ \ \ \ \ \ T_{\rm
photon} \gg T.
 \label{68}
\end{equation}
With the execution of inequalities (68) the charged particle may
be described by the classical Lagrangian
\begin{equation}
L = - M_0 c^2 \sqrt{1-v_0^2/c^2} - e\varphi + ev_0 \vec A
\label{69}
\end{equation}
where the velocity ${\vec v}_0$  describes the motion of the whole
particle complex that is formed around the particle in the space
tessellattice when the particle is created.

    If an incident photon flux is specified by a strong intensity
(for instance an intensive laser pulse), then a number of photons
will fall on the aforementioned particle complex. In this case one
should take into account the strong interaction between $\cal N$
external photons and the particle's inerton-photon cloud. The
theory of the anomalous photoelectric effect based on such
interaction has been constructed in paper [33].

\section*{\small 7. CONCLUDING REMARKS}

\hspace*{\parindent} The theory of the real physical space
developed in works [28-30] has allowed the construction of the
theory of the charge in the form of a peculiar quantum of fractal
deformations, or a chestnut created from a superparticle that is
the building block of the space. The total number of
superparticles of the space tessellattice that spreads and forms
all the universe defines the number of the charge needles
estimated by the value of ${{\mathfrak{N}}} \sim 10^{168}$. The
approach proposed has allowed the geometric interpretation of
major notions of electrodynamics such as the electric field vector
$\vec E$ and the vector potential $\vec A$ of the magnetic field.

    A particle and inerton-photon quasi-particles, which
accompany it, carry the inert mass, electric charge and spin.
Free, or real quasi-particles, irradiated by a particle should
carry at least some of these properties as well. Namely, a free
inerton that is radiated by a particle leaving its inerton cloud
must carries the mass. A free photon should carry the
electromagnetic polarization and, due to its location on the
inerton, the photon should carry also the mass (see also Ref.
[35]). These two quasi-particles move uniformly and rectilinearly
already on a microscopic scale (Fig. 7), though at the
submicroscopic consideration they travel by the relay mechanism
hopping from superparticle to superparticle. The particle cannot
move uniformly on a microscopic scale, its motion is adiabatic and
non-stationary: Owing to the interaction with surrounding
superparticles the particle moves with oscillating velocity
[19-22]. However, on the macroscopic scale that prevails the size
of the de Broglie wavelength $\lambda$  the behavior of the
particle resembles that of a classical object; and the behavior is
exactly classical on a scale larger than the particle's inerton
cloud amplitude $\Lambda$.
\begin{figure}
\begin{center}
\includegraphics[scale=1.04]{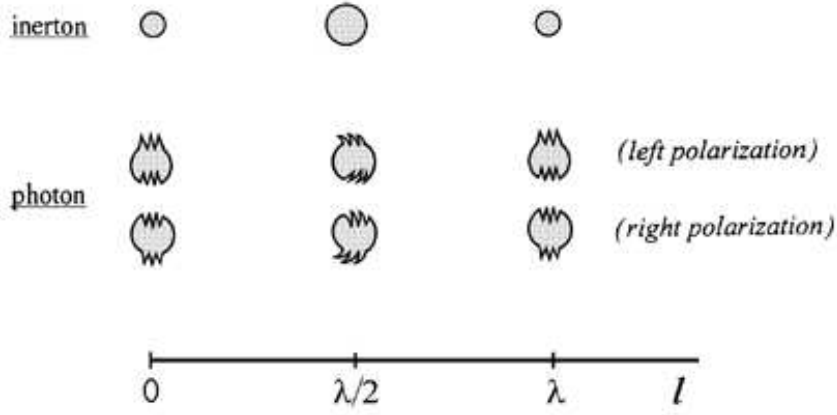}
\caption{\small Scheme of the motion of two basic quasi-particles
of the space. $(a)$  the motion of the free inerton; the volume of
the inerton oscillates along its path $l$ with the spatial period
$\lambda$. $(b)$  the motion of the free photon; the two possible
transverse polarizations of the photon are shown. The polarization
oscillates along the photon path $l$ with the spatial period
$\lambda$.}
 \label{Figure 7}
\end{center}
\end{figure}

The electric and magnetic polarizations are evolved separately, as
Eqs. (49) and (46) demonstrate, and at the maximum distance from
the particle, $\Lambda$, the electric polarization of the
inerton-photon cloud (caused by the field $\varphi$) drops to
zero, but the magnetic polarization reaches its maximum,
${\vec\alpha}_{\rm max}$. Since the mass of the inerton cloud
gradually decreases to zero at the distance $\Lambda$ from the
particle by Newton's law $1/r$ [24], we may anticipate that the
value of the electric polarization of the cloud should decrease by
the same law $1/r$, i.e. Coulomb's law.

In Ref. [35] and in the present work we have argued that the
photon polarization is kept on the surface of the corresponding
inerton and, because of that, the photon is characterized by the
mass as well. It must be emphasized once again that de Broglie
[36,37] was the first who conjectured that the photon was
characterizing by mass. Besides, the geometry of the charge
constructed clearly demonstrates the availability of an additional
mass introduced by the fractal deformations of the surface. Indeed
the total volume of the particle should consists of two terms: the
background volume $4\pi R^3_{\rm part}/3$ that corresponds to the
particle mass and the volume of all the needles
${\mathfrak{N}}V^{(0)}_{\rm needle}$ (see expression (28)), though
this correction seems should be neglected.

   The present model of the electric charge leads to the Maxwell
equations and hence the model perfectly agrees with classical
electrodynamics. However, it is only true in the case when the
frequency $\nu = E/h$ of proper oscillations of the particle along
its path satisfies the inequality $\nu > \nu_{\rm photon}$ where
$\nu_{\rm photon}$ is the frequency of a photon of an applied
electromagnetic field (note here $E= Mv_0^2/2$ where $M\equiv
M_0/\sqrt{1-v_0^2/c^2}$). With the opposite inequality, $\nu <
\nu_{\rm photon}$, effects of strong interaction should take
place, because the scattering of the photon by the inerton-photon
cloud of the particle becomes inelastic, i.e., an outside photon
penetrates into the cloud. The Compton scattering of light by the
particle occurs at the boundary of the particle's space
crystallite whose size is $\lambda_{{\kern 1pt} \rm C}=Mc/h$
[19-25].

The screening of the particle charge is restricted by the
crystallite and therefore the crystallite electric field must be
extraordinary powerful. That is why the field should strongly
polarize the crystallite creating virtual particle-antiparticle
pairs. Such effects fall within the study of high energy physics.
The theory of the real physical space and submicroscopic mechanics
are capable to contribute a lot to high energy physics clarifying
major notions of particle physics, hidden dynamics of particle
interactions and inner reasons of particle transformations.

The concept of submicroscopic fractality starting on the scale of
about $10^{-30}$ m is distributed to cosmic scale subdividing the
universe into ranges of intermediate levels (quarks, atoms,
molecules, human size, star systems, etc.) [29,30]. Besides, it
introduces charge topology in both the micro- and macroscopic
world. For instance, typical manifestations of fractal geometry, a
quantum of fractal deformations, on a macroscopic scale are a
chestnut, flower, crown of a tree, hedgehog and so on (positive
charges). Furthermore, if we determine the characteristic
fractality of the root of plant and the male sex as a positive
charge, we automatically should recognize that the opposite
fractality, namely, the stomach of a living entity and the female
sex is a typical negative charge.

It is interesting to emphasize that the submicroscopic fractality
of space would be an actual ground for the justification of
isomathematics that has been developing by Santilli [46] and his
followers for the strongly interacting systems, from hadrons to
molecules and molecular compounds (see, e.g. Ref. [47]).

Finally, let us turn to the epigraph to this paper, which was
taken from the recent book {\it Vedic Physics} by Dr. Raja Ram
Mohan Roy [48]. Roy analyzed the Vedic literature in detail
reading it as a physicist and he was not acquainted with the
author's study of the real space. His decoding of {\it The
\d{R}gveda}, which means 'the top of knowledge' in word-to-word
translation to modern languages, shows that this is a book about
the constitution of the real space, particle physics and
cosmology. Indeed: i) the epigraph above (Roy's decoding) exactly
corresponds to the contents of the present paper; ii) the three
steps of God Vi\d{s}nu (the universe) represent the space web by
Roy's decoding, which consists of indivisible cells, their
interface (i.e., the surface) and the observer space (i.e., the
aggregation of cells), which also in agreement with our theory.
Besides, all other details uncovered by Roy are in accord with
contemporary physics. Therefore it cannot be overestimated that
the physics pattern presented in the study above, including Refs.
[28-30], {\textit{\textbf{exactly}}} matches the Vedic picture of
the physical world, which has been unravelled in Roy's {\it Vedic
Physics} [48].

\section*{\small APPENDIX}

\hspace*{\parindent} Let the shape of the surface needle be
defined by the  figure of rotation (Fig. 8) of the function
 $$ y=h \cdot \bigl( 1-\frac {x^2}{r^2}
\bigr) \eqno (A1)
 $$
round the $OY$ axis. The area of surface of the needle is

\begin{figure}
\begin{center}
\includegraphics[scale=2.6]{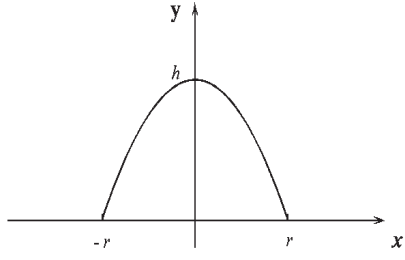}
\caption{\small Graphic display of the function (A1).}
\label{Figure 8}
\end{center}
\end{figure}

\noindent
 $$
\sigma_{\rm needle} = 2\pi \int\limits_0^h x (y){\kern 1pt} d l =
2\pi \int\limits_0^h r\sqrt{1-y/h} \ \sqrt{1+\frac {r^2}
{4h^2(1-y/h)}} \  d y
 $$
 $$ = \frac {4 \pi}{3} {\kern 1.51pt} r h \cdot \Bigl\{ \Bigl[ 1+ \Bigl( {\frac {r^2}
{2h}} \Bigr)^2 {\kern 1.5pt} \Bigr]^{3/2} - \Bigl( \frac {r}{2h}
\Bigr)^3  \Bigr\}.    \eqno (A2)
 $$
 In this case the surface defect is defined as
 $$
\Delta S (h,r)= {{\mathfrak{N}}} \cdot \bigl[ \sigma_{\rm needle}
(h,r) -\pi r^2 \bigr]. \eqno (A3)
 $$
 The derivatives necessary for the
analysis of the needle stability are the following:
 $$ \Delta
S^{{\kern 1pt}\prime}_r/{{\mathfrak{N}}}= \frac {2\pi}3 \frac
{r}{\varkappa} {\kern 1pt} (1+ \varkappa^{{\kern 1pt} 2})^{3/2} +
2\pi r \varkappa \sqrt{1+ \varkappa^{{\kern 1pt} 2}} - \frac
{8\pi}3 r \varkappa^{{\kern 1pt} 2} - 2\pi r; \eqno (A4)
 $$
 $$ \Delta
S^{{\kern 1pt} \prime}_h/{{\mathfrak{N}}}= \frac {4\pi}3 {\kern
1pt} r{\kern 1pt} (1+ \varkappa^{{\kern 1pt} 2})^{3/2} - 4\pi r
\varkappa^{{\kern 1pt} 2} \sqrt {1+ \varkappa^{{\kern 1pt} 2}} +
\frac {8\pi}3 r\varkappa^{{\kern 1pt} 3}; \eqno (A5)
 $$
 $$ \Delta S^{{\kern 1pt} \prime
\prime}_{rr}/{{\mathfrak{N}}}= 6\pi \varkappa {\kern 1pt}
\sqrt{1+\varkappa^{{\kern 1pt} 2}} + 2 \pi \frac
{\varkappa^{{\kern 1pt} 3}}{\sqrt{1+ \varkappa^{{\kern 1pt} 2}}} -
8\pi \varkappa^{{\kern 1pt} 2} - 2\pi; \eqno (A6)
 $$
 $$ \Delta S^{{\kern 1pt} \prime
\prime}_{hh}/{{\mathfrak{N}}}= 8\pi \varkappa^{{\kern 1pt} 3}
\sqrt{1+\varkappa^{{\kern 1pt} 2}} + 8\pi {\varkappa^{{\kern 1pt}
5}}{\sqrt{1+ \varkappa^{{\kern 1pt} 2}}} - 18\pi \varkappa^{{\kern
1pt} 4}; \eqno (A7)
 $$
 $$ \Delta S^{{\kern 1pt} \prime
\prime}_{rh}/{{\mathfrak{N}}}= \frac {4\pi}3 \bigl(
1+\varkappa^{{\kern 1pt} 2} \bigr)^{3/2} - 8\pi \varkappa^{{\kern
1pt} 2} \sqrt{1+ \varkappa^{{\kern 1pt} 2}} - 4\pi
{\varkappa^{{\kern 1pt} 4} \over \sqrt{1 + \varkappa^{{\kern 1pt}
2}}} + 8 \pi \varkappa^{{\kern 1pt} 3}; \eqno (A8)
 $$
 $$ \Delta S^{{\kern 1pt} \prime \prime}_{hr}/{{\mathfrak{N}}}=
{4\pi \over 3} \bigl( 1+\varkappa^{{\kern 1pt} 2} \bigr)^{3/2} -
8\pi \varkappa^{{\kern 1pt} 2} \sqrt{1+ \varkappa^{{\kern 1pt} 2}}
- 2\pi {\varkappa^{{\kern 1pt} 4}}{\sqrt{1+ \varkappa^{{\kern 1pt}
2}}} \eqno (A9)
 $$
where $\varkappa =r/2h$. Equations $\Delta S^{{\kern 1pt}
\prime}_h=0$ and $\Delta S^{{\kern 1pt} \prime}_r=0$ in the
explicit form are
 $$
 \frac 1{3\varkappa}(1+\varkappa^{{\kern 1pt} 2})^{3/2}=1
+\frac {\varkappa^{{\kern 1pt} 3}}3 - \varkappa \sqrt{1+
\varkappa^{{\kern 1pt} 2}}; \eqno (A10)
 $$
 $$ \frac 1{3\varkappa}(1+\varkappa^{{\kern 1pt} 2})^{3/2}= \varkappa
\sqrt{1+ \varkappa^{{\kern 1pt} 2}} - \frac 23 \varkappa^{{\kern
1pt} 2}. \eqno (A11)
 $$
Compatibility of Eqs. (A10) and (A11) gives the following equation
for $\varkappa$:
 $$ \varkappa^{{\kern 1pt} 3} - 2\sqrt{1+\varkappa^{{\kern 1pt} 2}} +1 =0.
\eqno (A12)
 $$
The solution to Eq. (A12) is
 $$ \varkappa
=\varkappa_{{\kern 1pt} 0} \simeq 1.323311. \eqno (A13)
 $$
Substituting $\varkappa_{{\kern 1pt} 0}$ into expressions
(A6)-(A9) we obtain
 $$ \Delta
S^{{\kern 1pt} \prime \prime}_{rr}(\varkappa_{{\kern 1pt} 0})/2\pi
{{\mathfrak{N}}}\simeq -1; \eqno (A14)
 $$
 $$ \Delta S^{{\kern 1pt} \prime
\prime}_{hh}(\varkappa_{{\kern 1pt} 0})/2\pi
{{\mathfrak{N}}}\simeq 10^{-6}; \eqno (A15)
 $$
 $$ \Delta S^{{\kern 1pt} \prime \prime}_{rh}(\varkappa_0)/2\pi
{\mathfrak{N}}\simeq 0.882208; \eqno (A16)
 $$
 $$ \Delta S^{{\kern 1pt} \prime
\prime}_{hr}(\varkappa_0)/2\pi {{\mathfrak{N}}}\simeq 3.1995296.
\eqno (A17)
 $$
By substituting the values of (A14)-(A17) into inequalities (12)
and (13) we get
 $$
 \frac {{{\rm Det}
\vert}_{\varkappa_0}} {2\pi {{\mathfrak{N}}}} = -2.82265 <0; \ \ \
\ \Delta S^{{\kern 1pt} \prime \prime}_{hh}(\varkappa_{{\kern 1pt}
0})>0, \ \ \ \ \Delta S^{{\kern 1pt} \prime \prime}
_{rr}(\varkappa_{{\kern 1pt} 0})<0. \eqno (A18)
 $$
Inequalities (A18) point out that the surface defect (A3) has no
extreme. Therefore the considered shape of the surface needle (A1)
is unstable and hence cannot form the charge state in a cell of
the tessellattice.

\end{document}